\documentclass[aip,apl, twocolumn, amssymb, amsmath, reprint]{revtex4-1}

\usepackage{graphicx}
\usepackage{dcolumn}
\usepackage{bm}
\usepackage{color}

\begin{document}
\title{Topological edge states in single- and multi-layer Bi$_{4}$Br$_{4}$ }

\author{Jin-Jian Zhou}
\affiliation{Institute of Physics, Chinese Academy of Sciences and Beijing
National Laboratory for Condensed Matter Physics, Beijing 100190, China }

\author{Wanxiang Feng}
\affiliation{School of Physics, Beijing Institute of Technology, Beijing 100081, China}

\author{Gui-Bin Liu}
\affiliation{School of Physics, Beijing Institute of Technology, Beijing 100081, China}

\author{Yugui Yao}
\email{ygyao@bit.edu.cn}
\affiliation{School of Physics, Beijing Institute of Technology, Beijing 100081, China}
\affiliation{Institute of Physics, Chinese Academy of Sciences and Beijing
National Laboratory for Condensed Matter Physics, Beijing 100190, China }

\date{\today}

\begin{abstract}
Topological edge states at the boundary of quantum spin Hall (QSH) insulators hold great promise for dissipationless electron transport. The device application of topological edge states has several critical requirements for the QSH insulator materials, e.g., large band gap, appropriate insulating substrates, and multiple conducting channels. In this paper, based on first-principle calculations, we show that Bi$_{4}$Br$_{4}$ is a suitable candidate. Single-layer Bi$_{4}$Br$_{4}$ was demonstrated to be QSH insulator with sizable gap recently. Here, we find that, in multilayer systems, both the band gaps and low-energy electronic structures are only slightly affected by the interlayer coupling. On the intrinsic insulating substrate of Bi$_{4}$Br$_{4}$, the single-layer Bi$_{4}$Br$_{4}$ well preserves its topological edge states. 
Moreover, at the boundary of multilayer Bi$_{4}$Br$_{4}$, the topological edge states stemming from different single-layers are weakly coupled , and can be fully decoupled via constructing a stair-stepped edge. The decoupled topological edge states are well suitable for multi-channel dissipationless transport.
\end{abstract}

\maketitle 

%\section{Introduction}
The hallmark of quantum spin Hall (QSH) insulators, also known as two-dimensional (2D) topological insulators, is the gapless helical edge states inside the bulk band gap\cite{Hasan2010,Qi2011}. Along a given edge of QSH insulator, a pair of edge states with opposite spins propagate in opposite directions, and they are topologically protected against backscattering from non-magnetic disorder. With this novel property, topological edge state promises its application to dissipationless transport, which has been demonstrated experimentally in HgTe/CdTe\cite{Koenig2007} and InAs/GaSb\cite{Knez2011} quantum wells. Despite the great promise, the device application of topological edge state has been hampered by the lack of  suitable materials that meet several critical requirements, e.g., large band gap for room temperature applications,  and multiple conducting channels for high signal-to-noise ratio.

Inspired by the discovery of graphene,  2D materials with atomic thickness have become an emerging playground for exploring novel physics. The QSH effect was firstly predicted in graphene\cite{Kane2005}, in which the band gap opened by spin-orbital couping (SOC) is extremely small\cite{Yao2007}.  
Subsequently, some honeycomb-like materials with heavier elements were proposed to be QSH insulators with experimental accessible gaps, such as silicene\cite{Liu2011} and Bi(111) bilayer\cite{Murakami2006}.
However, the lack of appropriate insulating substrates becomes another crucial issue. The QSH phase of 2D materials may be destroyed due to its interaction with substrates\cite{Chen2013}. Even the QSH phase survives, the hybridization between the topological edge states and the substrate's bulk states is disturbing\cite{Hirahara2011,Yang2012}. Topological edge modes of  Bi-bilayer on the surface of Bi single crystal were detected by STM recently\cite{Drozdov2014}, thanks to that the edge states of certain edge type are only slightly hybridized with bulk Bi. Yet the metallic surface of bulk Bi is inadequate for edge state transport. Other newly proposed QSH insulators with sizable band gaps, such as single-layer Bi$_{4}$Br$_{4}$\cite{Zhou2014} and transition metal dichalcogenides\cite{Qian2014}, may break this obstruction because their bulk crystals are insulators. 

In this paper, based on first-principles calculations, we study the effect of interlayer coupling on the electronic structures and edge states of multilayer Bi$_{4}$Br$_{4}$.  We find that the band gaps of multilayer Bi$_{4}$Br$_{4}$ hardly changes as the number of layers increasing, and the interlayer coupling has small impact on the low-energy electronic structures of multilayer Bi$_{4}$Br$_{4}$, which is attributed to the special orbital character of the band edges. 
Therefore, the surface of bulk Bi$_{4}$Br$_{4}$ can be an intrinsic insulating substrate for the SL Bi$_{4}$Br$_{4}$. With this substrate, the Fermi velocity of topological edge states is slightly reduced compared to the freestanding case. Moreover, at the boundary of multilayer Bi$_{4}$Br$_{4}$, the topological edge states stemming from different single-layers are weakly coupled, which can be further decoupled by constructing a stair-stepped edge. The decoupled topological edge states can serve as multiple conducting channels. Our results indicate that the Bi$_{4}$Br$_{4}$ is an excellent candidate for manufacturing multi-channel dissipathionless electron device. 
 
\begin{figure}
\includegraphics[width=8cm]{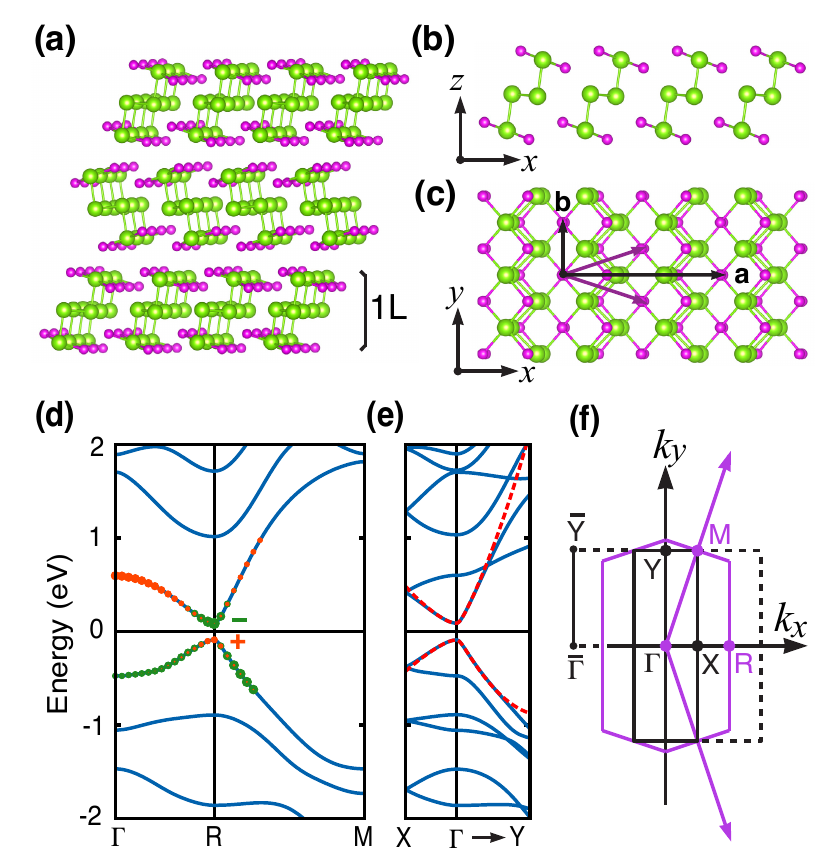}
\caption{ Side view of (a) multilayer and (b) SL Bi$_{4}$Br$_{4}$ with green (small purple) balls representing Bi (Br). Bi atoms in the middle of SL are labeled as Bi$_{in}$,  while these bonded with Br are labeled as Bi$_{ex}$. (c) Top view of SL Bi$_{4}$Br$_{4}$. The black (purple) vectors are lattice vectors of conventional (primitive) unit cell. (d) HSE06 band structure of SL Bi$_{4}$Br$_{4}$ with orbital projected character in the low-energy region. The size of orange and green circle denote the weight of Bi$_{ex}$-$p_{x}$ and Bi$_{in}$-$p_{x}$  orbital projections, respectively. (e) Comparison of band structures under conventional cell from HSE06 calculation (blue line) and the fitted $k\cdot p$ Hamiltonian (red dashed line). (f) the Brillouin Zones of primitive (purple) and conventional cell (black). When conventional cell is adopted, the Brillouin Zone of primitive cell is folded, e.g., the R point under primitive cell is folded to the $\Gamma$ point under conventional cell. }
\label{fig1}
\end{figure}

%\section{Methods}
First principle calculations are carried out using the projector augmented wave method~\cite{Blochl1994} as implemented in the Vienna \textit{ab initio} simulation package~\cite{Kresse1996}.  Both the Perdew-Burke-Ernzerof  generalized gradient approximation (GGA)\cite{Perdew1996} and the Heyd-Scuseria-Ernzerhof hybrid functional (HSE06)\cite{Heyd2006} are used for the exchange-correlation potential. 
The ionic position are relaxed employing the van der Waals (vdW) corrections~\cite{Dion2004,klime2011}. Maximally localized Wannier functions (MLWFs) for the $p$-orbitals of Bi and Br atoms are constructed using the \textsc{wannier90} code\cite{Mostofi2008}. In the HSE06 calculations, the surface electronic structures are calculated using the combination of  MLWFs and surface Green's function methods\cite{Sancho1985}.

\begin{figure}
\includegraphics[width=8cm]{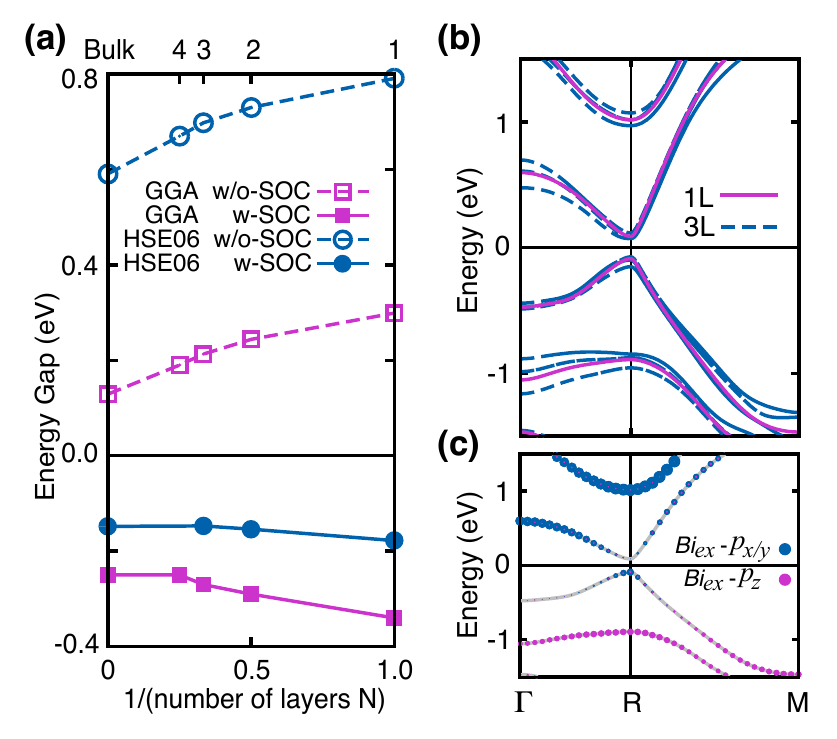}
\caption{ (a) The dependence of energy gap of few-layer Bi$_{4}$Br$_{4}$ on the number of layers calculated by using GGA and HSE06 potentials. (b) Comparison between HSE06 band structures of single-layer and triple-layer Bi$_{4}$Br$_{4}$. (c) HSE06 band structures of SL Bi$_{4}$Br$_{4}$ with orbital projected character. The size of  blue (purple) circle denotes the weights of Bi$_{ex}$-$p_{x/y}$ (Bi$_{ex}$-$p_{z}$) orbital projections. 
} \label{fig2}
\end{figure}

%\section{Result and Discussion}
Bi$_{4}$Br$_{4}$ has a layered structure\cite{Dikarev2001,Filatova2007}, as shown in Fig.~\ref{fig1}(a). Within each single-layer (SL), one Bi atomic layer is sandwiched by two Bi/Br atomic layers. The normal and mirror-reflected SLs are stacked alternatively along the $z$ direction with the interactions between adjacent SLs of weak vdW-type. A SL of Bi$_{4}$Br$_{4}$ has a thickness of $\sim$7 \AA. It can be regarded as a parallel arrangement of one-dimensional (1D) infinite molecule chains[Fig.~\ref{fig1}(b)]. From the top view shown in Fig.~\ref{fig1}(c), one can see that the SL structure belongs to the centered rectangular lattice, whose primitive unit cell is half size of its conventional cell. The conventional unit cell consists of two 1D chains, and the lattice constants $a$ and $b$ are 13.064 \AA\ and 4.338 \AA\ , respectively\cite{Dikarev2001}.

The SL Bi$_{4}$Br$_{4}$ has been predicted to be QSH insulator in our recent work\cite{Zhou2014}. In the absence of SOC, the top of valence band (TVB), dominated by Bi$_{in}$-$p_{x}$ orbital, has odd parity under inversion symmetry; while the bottom of conduction band (BCB), dominated by Bi$_{ex}$-$p_{x}$ orbital, has even parity\cite{Zhou2014}. After turning on SOC, as shown in Fig.~\ref{fig1}(d), both the orbital character and parity of the band edges are inverted at the R point due to the strong SOC of Bismuth. This band inversion result in non-trivial topological phase ($\mathbb{Z}_2=1$) in SL Bi$_{4}$Br$_{4}$. 

From SL to multilayer system, the significant electronic properties may be altered by interlayer coupling, e.g., the direct to indirect band gap transition between monolayer and multilayer MoS$_{2}$\cite{Mak2010}. To study the effect of interlayer couping on electronic structure of Bi$_{4}$Br$_{4}$, especially for the inverted band gap, we calculate the band structures of few-layer Bi$_{4}$Br$_{4}$ using both GGA and HSE06 potentials. The calculated band gaps of Bi$_{4}$Br$_{4}$ from SL  to bulk systems are shown in Fig.~\ref{fig2}(a). 
In the absence of SOC, the band gaps of HSE06 are obviously larger than those of GGA. This is because the GGA calculation usually underestimates the band gap\cite{Heyd2005}. In both GGA and HSE06 calculations, the band gap slightly decreases as the number of layers increasing, and the differences of band gaps are within $\sim$0.2 eV. When SOC is turned on, the band gaps of multilayer Bi$_{4}$Br$_{4}$ are inverted in the similar way as SL system. The inverted band gaps are presented with negative values in Fig.~\ref{fig2}(a). The dependence of band gap on the number of layers is further reduced. For HSE06 result, the band gap difference between SL and bulk systems is only within $\sim$30 meV, which is a very small value compared to other layered materials, e.g., black phosphorus ($\sim$0.7 eV)\cite{Liu2014}.

Apart from the band gaps, the low-energy electronic structures are also insensitive to the interlayer coupling.  Fig.~\ref{fig2}(b) shows the comparison between HSE06 band structures of  SL and triple-layer Bi$_{4}$Br$_{4}$. The band features are essentially the same for both systems. Each band in SL Bi$_{4}$Br$_{4}$ corresponds to three bands in triple-layer Bi$_{4}$Br$_{4}$, which are split due to the interlayer coupling. As can be seen in Fig.~\ref{fig2}(b), The band splitting around Fermi-level is rather small, e.g., the band splitting of  BCB (TVB) is about 40 (80) meV.

\begin{figure}
\includegraphics[width=8cm]{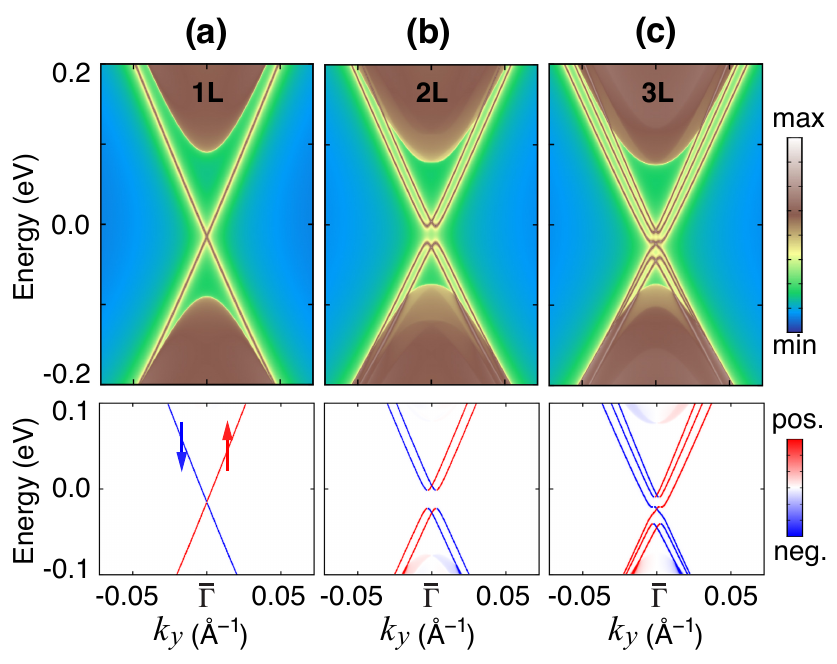}
\caption{ The edge density of states (first row) and spin polarization (second row) of (a) single-layer, (b) double-layer, and (c) triple-layer Bi$_{4}$Br$_{4}$. The Fermi level is set to the bulk band gap center. }\label{fig3}
\end{figure}

The small band splittings can be attributed to the special orbital characters of the band edges. When SLs are stacked together to form a multilayer structure, the states dominated by orbitals with larger interlayer hopping usually have larger band splitting in multilayer systems. Obviously, the out-of-plane $p_{z}$ orbital has larger interlayer hopping compared to the in-plane $p_{x/y}$ orbitals, and the orbitals from Bi$_{ex}$ have larger 
interlayer hopping compared to those from Bi$_{in}$. The band structure of  SL Bi$_{4}$Br$_{4}$ with orbital projected characters is plotted in Fig.~\ref{fig2}(c). The second valence bands are dominated by Bi$_{ex}$-$p_{z}$ orbital, thus have relatively large band splitting of a few hundred meV in triple-layer system [Fig.~\ref{fig2}(b)]. In contrast, the low-energy bands are dominated by the in-plane $p_{x/y}$ orbitals , mainly Bi$_{in}$-$p_{x}$ [Fig.~\ref{fig1}(d)], therefore they are less affected by the interlayer coupling. 

In the weak coupling limit, the multilayer system can be regarded as simple stacking of many isolated SLs which have topological edge states at the boundaries. We now focus on the evolution of these topological edges states when a weak interlayer coupling is introduced, such as the case in multilayer Bi$_{4}$Br$_{4}$. The edge electronic structures and spin polarizations for the SL, double-layer and triple-layer Bi$_{4}$Br$_{4}$ are shown in Fig.~\ref{fig3}.  
Since the coupling between adjacent 1D chains is much weaker than the intra-chain bonding\cite{Dikarev2001}, atomically sharp edges along the 1D chain axis ($y$-direction) without dangling bond can be stabilized. 
We construct such edges as semi-infinite systems, for which the surface Green's functions are calculated. 
The energy and momentum dependent density of states, that are extracted from the imaginary part of the surface Green's function, are used to analyze the edge electronic structures. For SL Bi$_{4}$Br$_{4}$ [Fig.~\ref{fig3}(a)], a single-Dirac-cone edges states linearly cross the bulk band gap. The Fermi velocity calculated by HSE06 is $\sim6.5\times 10^{5}$ m$\cdot$s$^{-1}$, which is larger than the GGA result\cite{Zhou2014}. For double-layer Bi$_{4}$Br$_{4}$ [Fig.~\ref{fig3}(b)], two pairs of topological edges states are weakly coupled, and a small gap of $\sim$20 meV is opened. The gapped edge states indicate topological trivial phase in double-layer Bi$_{4}$Br$_{4}$. For triple-layer Bi$_{4}$Br$_{4}$ [Fig.~\ref{fig3}(c)], three pairs of topological edges states are coupled. However,  there is one pair of edge states cross the band gap without gap opening, which indicates topological non-trivial phase in triple-layer Bi$_{4}$Br$_{4}$. The difference of topological phase is because that the multilayer Bi$_{4}$Br$_{4}$ with even (odd) number of layers has even (odd) times of band inversions. The even times of band inversions result in a trivial insulator with $\mathbb{Z}_{2}=0$, while odd times of band inversions result in a QSH insulator with $\mathbb{Z}_{2}=1$.

\begin{figure}
\includegraphics[width=8cm]{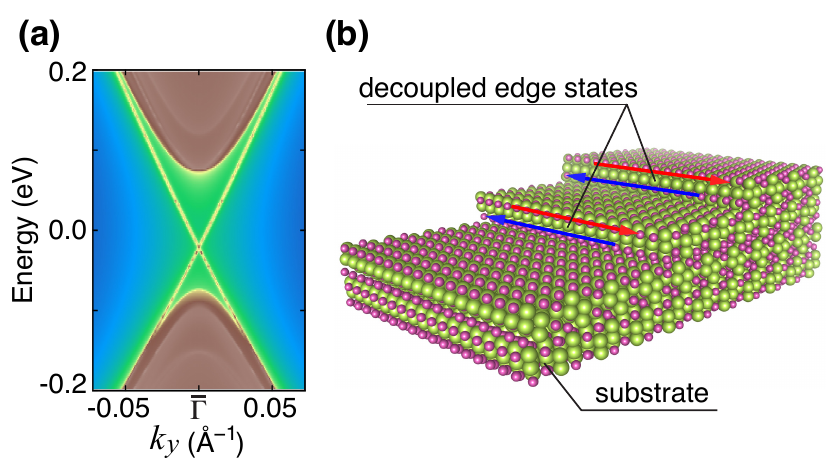}
\caption{ (a) The density of states of  Bi$_{4}$Br$_{4}$ with structures schematically plotted in (b).}\label{fig4}
\end{figure}

Since the topological edge states are weakly couped at the boundary of multilayer Bi$_{4}$Br$_{4}$,  they can be decoupled by constructing rough edge, such as a stair-stepped edge. We construct a two-layer Bi$_{4}$Br$_{4}$ film with stair-stepped edge supported by bulk Bi$_{4}$Br$_{4}$ surface, as plotted schematically in Fig.~\ref{fig4}(b). With the step width of  $\sim$5 nm, we calculate the energy spectrum of this system as shown in Fig.~\ref{fig4}(a). The edge states linearly cross the bulk gap without gap-opening. Compared to the freestanding double-layer one[Fig.~\ref{fig3}(b)], the topological edge states from the two different layers are fully decoupled at the stair-stepped edge, and the two single-Dirac-cones are degenerate. Another observation is that, due to the weak interaction with substrate, the Fermi velocity of the decoupled edge states ($\sim 5.6\times 10^{5}$ m$\cdot$s$^{-1}$) is a little smaller than that of the free standing SL system [Fig.~\ref{fig3}(a)]. As the effect of interlayer coupling, the decreased Fermi velocity is also observed in the edge states of triple-layer systems. 

To understand the essential physics, we develop a low-energy effective $k\cdot p$ Hamiltonian for SL Bi$_{4}$Br$_{4}$ by using the theory of invariants, from which the effective Hamiltonian for the topological edge states can be further derived. For the convenience to construct natural edges along the 1D chain axis, we adopt the conventional unit cell. Consequently the band edges are folded to the $\Gamma$-point, as illustrated in Fig.~\ref{fig1}(f). Both the BCB and TVB are double degenerate, and the two degenerate states are related by time reversal symmetry. Since the BCB (TVB) has odd (even) parity under inversion symmetry, we can denote these four states as $|-,\uparrow(\downarrow)\rangle$, $|+,\uparrow(\downarrow)\rangle$. By analyzing the inversion, mirror $\sigma_{h}$ and time reversal symmetry, we can write down  the low-energy effective Hamiltonian using the four states as basis (in the order of  $|+,\uparrow\rangle$, $|-,\downarrow\rangle$, $|+,\downarrow\rangle$,
$-|-,\uparrow\rangle$) 
\[
H_{\Gamma}(k)  =  \epsilon_{0}(k)  + 
\begin{pmatrix}
  M(k)                   &   A_{1}k_{x}       &   0                                         &   A_{2}k_{y}  \\
A_{1}^{*}k_{x}  &  -M(k)                  &  A_{2}k_{y}                        &   0                    \\
0                           & A_{2}^{*}k_{y}  &                 M(k)                     & -A_{1}^{*}k_{x}\\
A_{2}^{*}k_{y} & 0                            & -A_{1}k_{x}                        & -M(k)
\end{pmatrix}
\]
where $\epsilon_{0}(k)=C+D_{1}k_{x}^{2}+D_{2}k_{y}^{2}$, $M(k)=M_{0}-B_{1}k_{x}^{2}-B_{2}k_{y}^{2}$. $A_{1}$ and $A_{2}$ are complex parameters, while the others are real parameters. By fitting the energy spectrum of the Hamiltonian with the HSE06 band structure [see Fig.~\ref{fig1}(e)], we can determine these parameters as following: $C=0.0$ eV, $D_{1}=0.506$ eV$\cdot$\AA$^{2}$, $D_{2}=4.82$ eV$\cdot$\AA$^{2}$, $M_{0}=0.09$ eV, $B_{1}=3.86$ eV$\cdot$\AA$^{2}$, $B_{2}=0.0032$ eV$\cdot$\AA$^{2}$, $A_{1}=-1.81+0.046i $ eV$\cdot$\AA, $A_{2}=-4.15+0.141i$ eV$\cdot$\AA. 
The band inversion can be produced by the fact  of $M_{0}$, $B_{1}$, $B_{2} >  0$.  The form of the Hamiltonian  is different from that of HgTe quantum well\cite{Bernevig2006}, but similar to that of Bi$_{2}$Se$_{3}$\cite{Zhang2009}. In a similar way as for Bi$_{2}$Se$_{3}$, we can derive the effective Hamiltonian for the topological edge states. 
\[
H_{edge}(k_{y}) = \left|A_{2}\right| k_{y} \sigma_{x}
\]
With the fitted value of $A_{2}$, the Fermi velocity of the topological edge states is given by $\frac{\left|A_{2}\right|}{\hbar} \simeq 6.3\times 10^{5}$ m$\cdot$s$^{-1}$, which is consistent with the HSE06 result [Fig.~\ref{fig3}(a)]. The value of $A_{2}$ is modified when SL system is supported by Bi$_{4}$Br$_{4}$ surface. 
For the edge states of multilayer Bi$_{4}$Br$_{4}$, the effective Hamiltonian can be simply written by introducing coupling terms between $H_{edge}(k_{y})$ of different SLs. 

%\section{Conclusion}
In summary, first-principle calculations demonstrate that both the band gaps and low-energy electronic structures of multilayer Bi$_{4}$Br$_{4}$ are little affected by the interlayer coupling. When SL Bi$_{4}$Br$_{4}$ is supported by the surface of bulk Bi$_{4}$Br$_{4}$, its topological edge states well survive except for a reduced Fermi velocity. Moreover, at the stair-stepped edge of multilayer Bi$_{4}$Br$_{4}$, the topological edge states from different SLs are fully decoupled. 
Our results indicate nano-fabrication on the cleaved surface of  layered Bi$_{4}$Br$_{4}$ single crystal is adequate to realize multiple dissipationless conducting channels\cite{Sabater2013,Drozdov2014}, hence Bi$_{4}$Br$_{4}$ is an excellent platform for manufacturing QSH-based devices. 
\\

This work was supported by the MOST Project of China (Nos.~2014CB920903, 2013CB921903, 2011CBA00100), the NSF of China (Nos.~11174337, 11225418, 11374033, 11304014) and the SRFDPHE of China (No.~20121101110046, 20131101120052).

\bibliography{refs}

%merlin.mbs aipnum4-1.bst 2010-07-25 4.21a (PWD, AO, DPC) hacked
%Control: key (0)
%Control: author (8) initials jnrlst
%Control: editor formatted (1) identically to author
%Control: production of article title (-1) disabled
%Control: page (0) single
%Control: year (1) truncated
%Control: production of eprint (0) enabled
\begin{thebibliography}{30}%
\makeatletter
\providecommand \@ifxundefined [1]{%
 \@ifx{#1\undefined}
}%
\providecommand \@ifnum [1]{%
 \ifnum #1\expandafter \@firstoftwo
 \else \expandafter \@secondoftwo
 \fi
}%
\providecommand \@ifx [1]{%
 \ifx #1\expandafter \@firstoftwo
 \else \expandafter \@secondoftwo
 \fi
}%
\providecommand \natexlab [1]{#1}%
\providecommand \enquote  [1]{``#1''}%
\providecommand \bibnamefont  [1]{#1}%
\providecommand \bibfnamefont [1]{#1}%
\providecommand \citenamefont [1]{#1}%
\providecommand \href@noop [0]{\@secondoftwo}%
\providecommand \href [0]{\begingroup \@sanitize@url \@href}%
\providecommand \@href[1]{\@@startlink{#1}\@@href}%
\providecommand \@@href[1]{\endgroup#1\@@endlink}%
\providecommand \@sanitize@url [0]{\catcode `\\12\catcode `\$12\catcode
  `\&12\catcode `\#12\catcode `\^12\catcode `\_12\catcode `\%12\relax}%
\providecommand \@@startlink[1]{}%
\providecommand \@@endlink[0]{}%
\providecommand \url  [0]{\begingroup\@sanitize@url \@url }%
\providecommand \@url [1]{\endgroup\@href {#1}{\urlprefix }}%
\providecommand \urlprefix  [0]{URL }%
\providecommand \Eprint [0]{\href }%
\providecommand \doibase [0]{http://dx.doi.org/}%
\providecommand \selectlanguage [0]{\@gobble}%
\providecommand \bibinfo  [0]{\@secondoftwo}%
\providecommand \bibfield  [0]{\@secondoftwo}%
\providecommand \translation [1]{[#1]}%
\providecommand \BibitemOpen [0]{}%
\providecommand \bibitemStop [0]{}%
\providecommand \bibitemNoStop [0]{.\EOS\space}%
\providecommand \EOS [0]{\spacefactor3000\relax}%
\providecommand \BibitemShut  [1]{\csname bibitem#1\endcsname}%
\let\auto@bib@innerbib\@empty
%</preamble>
\bibitem [{\citenamefont {Hasan}\ and\ \citenamefont {Kane}(2010)}]{Hasan2010}%
  \BibitemOpen
  \bibfield  {author} {\bibinfo {author} {\bibfnamefont {M.~Z.}\ \bibnamefont
  {Hasan}}\ and\ \bibinfo {author} {\bibfnamefont {C.~L.}\ \bibnamefont
  {Kane}},\ }\href {\doibase 10.1103/RevModPhys.82.3045} {\bibfield  {journal}
  {\bibinfo  {journal} {Rev. Mod. Phys.}\ }\textbf {\bibinfo {volume} {82}},\
  \bibinfo {pages} {3045} (\bibinfo {year} {2010})}\BibitemShut {NoStop}%
\bibitem [{\citenamefont {Qi}\ and\ \citenamefont {Zhang}(2011)}]{Qi2011}%
  \BibitemOpen
  \bibfield  {author} {\bibinfo {author} {\bibfnamefont {X.-L.}\ \bibnamefont
  {Qi}}\ and\ \bibinfo {author} {\bibfnamefont {S.-C.}\ \bibnamefont {Zhang}},\
  }\href {\doibase 10.1103/RevModPhys.83.1057} {\bibfield  {journal} {\bibinfo
  {journal} {Rev. Mod. Phys.}\ }\textbf {\bibinfo {volume} {83}},\ \bibinfo
  {pages} {1057} (\bibinfo {year} {2011})}\BibitemShut {NoStop}%
\bibitem [{\citenamefont {K{\" o}nig}\ \emph {et~al.}(2007)\citenamefont {K{\"
  o}nig}, \citenamefont {Wiedmann}, \citenamefont {Brüne}, \citenamefont
  {Roth}, \citenamefont {Buhmann}, \citenamefont {Molenkamp}, \citenamefont
  {Qi},\ and\ \citenamefont {Zhang}}]{Koenig2007}%
  \BibitemOpen
  \bibfield  {author} {\bibinfo {author} {\bibfnamefont {M.}~\bibnamefont {K{\"
  o}nig}}, \bibinfo {author} {\bibfnamefont {S.}~\bibnamefont {Wiedmann}},
  \bibinfo {author} {\bibfnamefont {C.}~\bibnamefont {Brüne}}, \bibinfo
  {author} {\bibfnamefont {A.}~\bibnamefont {Roth}}, \bibinfo {author}
  {\bibfnamefont {H.}~\bibnamefont {Buhmann}}, \bibinfo {author} {\bibfnamefont
  {L.~W.}\ \bibnamefont {Molenkamp}}, \bibinfo {author} {\bibfnamefont {X.-L.}\
  \bibnamefont {Qi}}, \ and\ \bibinfo {author} {\bibfnamefont {S.-C.}\
  \bibnamefont {Zhang}},\ }\href {\doibase 10.1126/science.1148047} {\bibfield
  {journal} {\bibinfo  {journal} {Science}\ }\textbf {\bibinfo {volume}
  {318}},\ \bibinfo {pages} {766} (\bibinfo {year} {2007})}\BibitemShut
  {NoStop}%
\bibitem [{\citenamefont {Knez}, \citenamefont {Du},\ and\ \citenamefont
  {Sullivan}(2011)}]{Knez2011}%
  \BibitemOpen
  \bibfield  {author} {\bibinfo {author} {\bibfnamefont {I.}~\bibnamefont
  {Knez}}, \bibinfo {author} {\bibfnamefont {R.-R.}\ \bibnamefont {Du}}, \ and\
  \bibinfo {author} {\bibfnamefont {G.}~\bibnamefont {Sullivan}},\ }\href
  {\doibase 10.1103/PhysRevLett.107.136603} {\bibfield  {journal} {\bibinfo
  {journal} {Phys. Rev. Lett.}\ }\textbf {\bibinfo {volume} {107}},\ \bibinfo
  {pages} {136603} (\bibinfo {year} {2011})}\BibitemShut {NoStop}%
\bibitem [{\citenamefont {Kane}\ and\ \citenamefont {Mele}(2005)}]{Kane2005}%
  \BibitemOpen
  \bibfield  {author} {\bibinfo {author} {\bibfnamefont {C.~L.}\ \bibnamefont
  {Kane}}\ and\ \bibinfo {author} {\bibfnamefont {E.~J.}\ \bibnamefont
  {Mele}},\ }\href {\doibase 10.1103/PhysRevLett.95.226801} {\bibfield
  {journal} {\bibinfo  {journal} {Phys. Rev. Lett.}\ }\textbf {\bibinfo
  {volume} {95}},\ \bibinfo {pages} {226801} (\bibinfo {year}
  {2005})}\BibitemShut {NoStop}%
\bibitem [{\citenamefont {Yao}\ \emph {et~al.}(2007)\citenamefont {Yao},
  \citenamefont {Ye}, \citenamefont {Qi}, \citenamefont {Zhang},\ and\
  \citenamefont {Fang}}]{Yao2007}%
  \BibitemOpen
  \bibfield  {author} {\bibinfo {author} {\bibfnamefont {Y.}~\bibnamefont
  {Yao}}, \bibinfo {author} {\bibfnamefont {F.}~\bibnamefont {Ye}}, \bibinfo
  {author} {\bibfnamefont {X.-L.}\ \bibnamefont {Qi}}, \bibinfo {author}
  {\bibfnamefont {S.-C.}\ \bibnamefont {Zhang}}, \ and\ \bibinfo {author}
  {\bibfnamefont {Z.}~\bibnamefont {Fang}},\ }\href {\doibase
  10.1103/PhysRevB.75.041401} {\bibfield  {journal} {\bibinfo  {journal} {Phys.
  Rev. B}\ }\textbf {\bibinfo {volume} {75}},\ \bibinfo {pages} {041401}
  (\bibinfo {year} {2007})}\BibitemShut {NoStop}%
\bibitem [{\citenamefont {Liu}, \citenamefont {Feng},\ and\ \citenamefont
  {Yao}(2011)}]{Liu2011}%
  \BibitemOpen
  \bibfield  {author} {\bibinfo {author} {\bibfnamefont {C.-C.}\ \bibnamefont
  {Liu}}, \bibinfo {author} {\bibfnamefont {W.}~\bibnamefont {Feng}}, \ and\
  \bibinfo {author} {\bibfnamefont {Y.}~\bibnamefont {Yao}},\ }\href {\doibase
  10.1103/PhysRevLett.107.076802} {\bibfield  {journal} {\bibinfo  {journal}
  {Phys. Rev. Lett.}\ }\textbf {\bibinfo {volume} {107}},\ \bibinfo {pages}
  {076802} (\bibinfo {year} {2011})}\BibitemShut {NoStop}%
\bibitem [{\citenamefont {Murakami}(2006)}]{Murakami2006}%
  \BibitemOpen
  \bibfield  {author} {\bibinfo {author} {\bibfnamefont {S.}~\bibnamefont
  {Murakami}},\ }\href {\doibase 10.1103/PhysRevLett.97.236805} {\bibfield
  {journal} {\bibinfo  {journal} {Phys. Rev. Lett.}\ }\textbf {\bibinfo
  {volume} {97}},\ \bibinfo {pages} {236805} (\bibinfo {year}
  {2006})}\BibitemShut {NoStop}%
\bibitem [{\citenamefont {Chen}\ \emph {et~al.}(2013)\citenamefont {Chen},
  \citenamefont {Li}, \citenamefont {Feng}, \citenamefont {Ding}, \citenamefont
  {Qiu}, \citenamefont {Cheng}, \citenamefont {Wu},\ and\ \citenamefont
  {Meng}}]{Chen2013}%
  \BibitemOpen
  \bibfield  {author} {\bibinfo {author} {\bibfnamefont {L.}~\bibnamefont
  {Chen}}, \bibinfo {author} {\bibfnamefont {H.}~\bibnamefont {Li}}, \bibinfo
  {author} {\bibfnamefont {B.}~\bibnamefont {Feng}}, \bibinfo {author}
  {\bibfnamefont {Z.}~\bibnamefont {Ding}}, \bibinfo {author} {\bibfnamefont
  {J.}~\bibnamefont {Qiu}}, \bibinfo {author} {\bibfnamefont {P.}~\bibnamefont
  {Cheng}}, \bibinfo {author} {\bibfnamefont {K.}~\bibnamefont {Wu}}, \ and\
  \bibinfo {author} {\bibfnamefont {S.}~\bibnamefont {Meng}},\ }\href {\doibase
  10.1103/PhysRevLett.110.085504} {\bibfield  {journal} {\bibinfo  {journal}
  {Phys. Rev. Lett.}\ }\textbf {\bibinfo {volume} {110}},\ \bibinfo {pages}
  {085504} (\bibinfo {year} {2013})}\BibitemShut {NoStop}%
\bibitem [{\citenamefont {Hirahara}\ \emph {et~al.}(2011)\citenamefont
  {Hirahara}, \citenamefont {Bihlmayer}, \citenamefont {Sakamoto},
  \citenamefont {Yamada}, \citenamefont {Miyazaki}, \citenamefont {Kimura},
  \citenamefont {Bl{\" u}gel},\ and\ \citenamefont {Hasegawa}}]{Hirahara2011}%
  \BibitemOpen
  \bibfield  {author} {\bibinfo {author} {\bibfnamefont {T.}~\bibnamefont
  {Hirahara}}, \bibinfo {author} {\bibfnamefont {G.}~\bibnamefont {Bihlmayer}},
  \bibinfo {author} {\bibfnamefont {Y.}~\bibnamefont {Sakamoto}}, \bibinfo
  {author} {\bibfnamefont {M.}~\bibnamefont {Yamada}}, \bibinfo {author}
  {\bibfnamefont {H.}~\bibnamefont {Miyazaki}}, \bibinfo {author}
  {\bibfnamefont {S.-i.}\ \bibnamefont {Kimura}}, \bibinfo {author}
  {\bibfnamefont {S.}~\bibnamefont {Bl{\" u}gel}}, \ and\ \bibinfo {author}
  {\bibfnamefont {S.}~\bibnamefont {Hasegawa}},\ }\href {\doibase
  10.1103/PhysRevLett.107.166801} {\bibfield  {journal} {\bibinfo  {journal}
  {Phys. Rev. Lett.}\ }\textbf {\bibinfo {volume} {107}},\ \bibinfo {pages}
  {166801} (\bibinfo {year} {2011})}\BibitemShut {NoStop}%
\bibitem [{\citenamefont {Yang}\ \emph {et~al.}(2012)\citenamefont {Yang},
  \citenamefont {Miao}, \citenamefont {Wang}, \citenamefont {Yao},
  \citenamefont {Zhu}, \citenamefont {Song}, \citenamefont {Wang},
  \citenamefont {Xu}, \citenamefont {Fedorov}, \citenamefont {Sun},
  \citenamefont {Zhang}, \citenamefont {Liu}, \citenamefont {Liu},
  \citenamefont {Qian}, \citenamefont {Gao},\ and\ \citenamefont
  {Jia}}]{Yang2012}%
  \BibitemOpen
  \bibfield  {author} {\bibinfo {author} {\bibfnamefont {F.}~\bibnamefont
  {Yang}}, \bibinfo {author} {\bibfnamefont {L.}~\bibnamefont {Miao}}, \bibinfo
  {author} {\bibfnamefont {Z.~F.}\ \bibnamefont {Wang}}, \bibinfo {author}
  {\bibfnamefont {M.-Y.}\ \bibnamefont {Yao}}, \bibinfo {author} {\bibfnamefont
  {F.}~\bibnamefont {Zhu}}, \bibinfo {author} {\bibfnamefont {Y.~R.}\
  \bibnamefont {Song}}, \bibinfo {author} {\bibfnamefont {M.-X.}\ \bibnamefont
  {Wang}}, \bibinfo {author} {\bibfnamefont {J.-P.}\ \bibnamefont {Xu}},
  \bibinfo {author} {\bibfnamefont {A.~V.}\ \bibnamefont {Fedorov}}, \bibinfo
  {author} {\bibfnamefont {Z.}~\bibnamefont {Sun}}, \bibinfo {author}
  {\bibfnamefont {G.~B.}\ \bibnamefont {Zhang}}, \bibinfo {author}
  {\bibfnamefont {C.}~\bibnamefont {Liu}}, \bibinfo {author} {\bibfnamefont
  {F.}~\bibnamefont {Liu}}, \bibinfo {author} {\bibfnamefont {D.}~\bibnamefont
  {Qian}}, \bibinfo {author} {\bibfnamefont {C.~L.}\ \bibnamefont {Gao}}, \
  and\ \bibinfo {author} {\bibfnamefont {J.-F.}\ \bibnamefont {Jia}},\ }\href
  {\doibase 10.1103/PhysRevLett.109.016801} {\bibfield  {journal} {\bibinfo
  {journal} {Phys. Rev. Lett.}\ }\textbf {\bibinfo {volume} {109}},\ \bibinfo
  {pages} {016801} (\bibinfo {year} {2012})}\BibitemShut {NoStop}%
\bibitem [{\citenamefont {Drozdov}\ \emph {et~al.}(2014)\citenamefont
  {Drozdov}, \citenamefont {Alexandradinata}, \citenamefont {Jeon},
  \citenamefont {Nadj-Perge}, \citenamefont {Ji}, \citenamefont {Cava},
  \citenamefont {Andrei~Bernevig},\ and\ \citenamefont
  {Yazdani}}]{Drozdov2014}%
  \BibitemOpen
  \bibfield  {author} {\bibinfo {author} {\bibfnamefont {I.~K.}\ \bibnamefont
  {Drozdov}}, \bibinfo {author} {\bibfnamefont {A.}~\bibnamefont
  {Alexandradinata}}, \bibinfo {author} {\bibfnamefont {S.}~\bibnamefont
  {Jeon}}, \bibinfo {author} {\bibfnamefont {S.}~\bibnamefont {Nadj-Perge}},
  \bibinfo {author} {\bibfnamefont {H.}~\bibnamefont {Ji}}, \bibinfo {author}
  {\bibfnamefont {R.~J.}\ \bibnamefont {Cava}}, \bibinfo {author}
  {\bibfnamefont {B.}~\bibnamefont {Andrei~Bernevig}}, \ and\ \bibinfo {author}
  {\bibfnamefont {A.}~\bibnamefont {Yazdani}},\ }\href {\doibase
  doi:10.1038/nphys3048} {\bibfield  {journal} {\bibinfo  {journal} {Nat Phys}\
  ,\ } (\bibinfo {year} {2014})},\ \Eprint
  {http://arxiv.org/abs/DOI:10.1038/nphys3048} {DOI:10.1038/nphys3048}
  \BibitemShut {NoStop}%
\bibitem [{\citenamefont {Zhou}\ \emph {et~al.}(2014)\citenamefont {Zhou},
  \citenamefont {Feng}, \citenamefont {Liu}, \citenamefont {Guan},\ and\
  \citenamefont {Yao}}]{Zhou2014}%
  \BibitemOpen
  \bibfield  {author} {\bibinfo {author} {\bibfnamefont {J.-J.}\ \bibnamefont
  {Zhou}}, \bibinfo {author} {\bibfnamefont {W.}~\bibnamefont {Feng}}, \bibinfo
  {author} {\bibfnamefont {C.-C.}\ \bibnamefont {Liu}}, \bibinfo {author}
  {\bibfnamefont {S.}~\bibnamefont {Guan}}, \ and\ \bibinfo {author}
  {\bibfnamefont {Y.}~\bibnamefont {Yao}},\ }\href {\doibase 10.1021/nl501907g}
  {\bibfield  {journal} {\bibinfo  {journal} {Nano Letters}\ }\textbf {\bibinfo
  {volume} {14}},\ \bibinfo {pages} {4767} (\bibinfo {year}
  {2014})}\BibitemShut {NoStop}%
\bibitem [{\citenamefont {Qian}\ \emph {et~al.}(2014)\citenamefont {Qian},
  \citenamefont {Liu}, \citenamefont {Fu},\ and\ \citenamefont
  {Li}}]{Qian2014}%
  \BibitemOpen
  \bibfield  {author} {\bibinfo {author} {\bibfnamefont {X.}~\bibnamefont
  {Qian}}, \bibinfo {author} {\bibfnamefont {J.}~\bibnamefont {Liu}}, \bibinfo
  {author} {\bibfnamefont {L.}~\bibnamefont {Fu}}, \ and\ \bibinfo {author}
  {\bibfnamefont {J.}~\bibnamefont {Li}},\ }\href
  {http://arxiv.org/abs/1406.2749} {\bibfield  {journal} {\bibinfo  {journal}
  {{arXiv}:1406.2749 [cond-mat]}\ } (\bibinfo {year} {2014})}\BibitemShut
  {NoStop}%
\bibitem [{\citenamefont {Bl{\" o}chl}(1994)}]{Blochl1994}%
  \BibitemOpen
  \bibfield  {author} {\bibinfo {author} {\bibfnamefont {P.~E.}\ \bibnamefont
  {Bl{\" o}chl}},\ }\href {\doibase 10.1103/PhysRevB.50.17953} {\bibfield
  {journal} {\bibinfo  {journal} {Phys. Rev. B}\ }\textbf {\bibinfo {volume}
  {50}},\ \bibinfo {pages} {17953} (\bibinfo {year} {1994})}\BibitemShut
  {NoStop}%
\bibitem [{\citenamefont {Kresse}\ and\ \citenamefont {Furthm{\"
  u}ller}(1996)}]{Kresse1996}%
  \BibitemOpen
  \bibfield  {author} {\bibinfo {author} {\bibfnamefont {G.}~\bibnamefont
  {Kresse}}\ and\ \bibinfo {author} {\bibfnamefont {J.}~\bibnamefont {Furthm{\"
  u}ller}},\ }\href {\doibase 10.1103/PhysRevB.54.11169} {\bibfield  {journal}
  {\bibinfo  {journal} {Phys. Rev. B}\ }\textbf {\bibinfo {volume} {54}},\
  \bibinfo {pages} {11169} (\bibinfo {year} {1996})}\BibitemShut {NoStop}%
\bibitem [{\citenamefont {Perdew}, \citenamefont {Burke},\ and\ \citenamefont
  {Ernzerhof}(1996)}]{Perdew1996}%
  \BibitemOpen
  \bibfield  {author} {\bibinfo {author} {\bibfnamefont {J.~P.}\ \bibnamefont
  {Perdew}}, \bibinfo {author} {\bibfnamefont {K.}~\bibnamefont {Burke}}, \
  and\ \bibinfo {author} {\bibfnamefont {M.}~\bibnamefont {Ernzerhof}},\ }\href
  {\doibase 10.1103/PhysRevLett.77.3865} {\bibfield  {journal} {\bibinfo
  {journal} {Phys. Rev. Lett.}\ }\textbf {\bibinfo {volume} {77}},\ \bibinfo
  {pages} {3865} (\bibinfo {year} {1996})}\BibitemShut {NoStop}%
\bibitem [{\citenamefont {Heyd}, \citenamefont {Scuseria},\ and\ \citenamefont
  {Ernzerhof}(2006)}]{Heyd2006}%
  \BibitemOpen
  \bibfield  {author} {\bibinfo {author} {\bibfnamefont {J.}~\bibnamefont
  {Heyd}}, \bibinfo {author} {\bibfnamefont {G.~E.}\ \bibnamefont {Scuseria}},
  \ and\ \bibinfo {author} {\bibfnamefont {M.}~\bibnamefont {Ernzerhof}},\
  }\href {\doibase 10.1063/1.2204597} {\bibfield  {journal} {\bibinfo
  {journal} {J. Chem. Phys.}\ }\textbf {\bibinfo {volume} {124}},\ \bibinfo
  {pages} {219906} (\bibinfo {year} {2006})}\BibitemShut {NoStop}%
\bibitem [{\citenamefont {Dion}\ \emph {et~al.}(2004)\citenamefont {Dion},
  \citenamefont {Rydberg}, \citenamefont {Schr\"oder}, \citenamefont
  {Langreth},\ and\ \citenamefont {Lundqvist}}]{Dion2004}%
  \BibitemOpen
  \bibfield  {author} {\bibinfo {author} {\bibfnamefont {M.}~\bibnamefont
  {Dion}}, \bibinfo {author} {\bibfnamefont {H.}~\bibnamefont {Rydberg}},
  \bibinfo {author} {\bibfnamefont {E.}~\bibnamefont {Schr\"oder}}, \bibinfo
  {author} {\bibfnamefont {D.~C.}\ \bibnamefont {Langreth}}, \ and\ \bibinfo
  {author} {\bibfnamefont {B.~I.}\ \bibnamefont {Lundqvist}},\ }\href {\doibase
  10.1103/PhysRevLett.92.246401} {\bibfield  {journal} {\bibinfo  {journal}
  {Phys. Rev. Lett.}\ }\textbf {\bibinfo {volume} {92}},\ \bibinfo {pages}
  {246401} (\bibinfo {year} {2004})}\BibitemShut {NoStop}%
\bibitem [{\citenamefont {Klime\ifmmode~\check{s}\else \v{s}\fi{}},
  \citenamefont {Bowler},\ and\ \citenamefont {Michaelides}(2011)}]{klime2011}%
  \BibitemOpen
  \bibfield  {author} {\bibinfo {author} {\bibfnamefont {J.}~\bibnamefont
  {Klime\ifmmode~\check{s}\else \v{s}\fi{}}}, \bibinfo {author} {\bibfnamefont
  {D.~R.}\ \bibnamefont {Bowler}}, \ and\ \bibinfo {author} {\bibfnamefont
  {A.}~\bibnamefont {Michaelides}},\ }\href {\doibase
  10.1103/PhysRevB.83.195131} {\bibfield  {journal} {\bibinfo  {journal} {Phys.
  Rev. B}\ }\textbf {\bibinfo {volume} {83}},\ \bibinfo {pages} {195131}
  (\bibinfo {year} {2011})}\BibitemShut {NoStop}%
\bibitem [{\citenamefont {Mostofi}\ \emph {et~al.}(2008)\citenamefont
  {Mostofi}, \citenamefont {Yates}, \citenamefont {Lee}, \citenamefont {Souza},
  \citenamefont {Vanderbilt},\ and\ \citenamefont {Marzari}}]{Mostofi2008}%
  \BibitemOpen
  \bibfield  {author} {\bibinfo {author} {\bibfnamefont {A.~A.}\ \bibnamefont
  {Mostofi}}, \bibinfo {author} {\bibfnamefont {J.~R.}\ \bibnamefont {Yates}},
  \bibinfo {author} {\bibfnamefont {Y.-S.}\ \bibnamefont {Lee}}, \bibinfo
  {author} {\bibfnamefont {I.}~\bibnamefont {Souza}}, \bibinfo {author}
  {\bibfnamefont {D.}~\bibnamefont {Vanderbilt}}, \ and\ \bibinfo {author}
  {\bibfnamefont {N.}~\bibnamefont {Marzari}},\ }\href {\doibase DOI:
  10.1016/j.cpc.2007.11.016} {\bibfield  {journal} {\bibinfo  {journal}
  {Comput. Phys. Commun.}\ }\textbf {\bibinfo {volume} {178}},\ \bibinfo
  {pages} {685 } (\bibinfo {year} {2008})}\BibitemShut {NoStop}%
\bibitem [{\citenamefont {Sancho}\ \emph {et~al.}(1985)\citenamefont {Sancho},
  \citenamefont {Sancho}, \citenamefont {Sancho},\ and\ \citenamefont
  {Rubio}}]{Sancho1985}%
  \BibitemOpen
  \bibfield  {author} {\bibinfo {author} {\bibfnamefont {M.~P.~L.}\
  \bibnamefont {Sancho}}, \bibinfo {author} {\bibfnamefont {J.~M.~L.}\
  \bibnamefont {Sancho}}, \bibinfo {author} {\bibfnamefont {J.~M.~L.}\
  \bibnamefont {Sancho}}, \ and\ \bibinfo {author} {\bibfnamefont
  {J.}~\bibnamefont {Rubio}},\ }\href
  {http://stacks.iop.org/0305-4608/15/i=4/a=009} {\bibfield  {journal}
  {\bibinfo  {journal} {Journal of Physics F: Metal Physics}\ }\textbf
  {\bibinfo {volume} {15}},\ \bibinfo {pages} {851} (\bibinfo {year}
  {1985})}\BibitemShut {NoStop}%
\bibitem [{\citenamefont {Dikarev}, \citenamefont {Popovkin},\ and\
  \citenamefont {Shevelkov}(2001)}]{Dikarev2001}%
  \BibitemOpen
  \bibfield  {author} {\bibinfo {author} {\bibfnamefont {E.~V.}\ \bibnamefont
  {Dikarev}}, \bibinfo {author} {\bibfnamefont {B.~A.}\ \bibnamefont
  {Popovkin}}, \ and\ \bibinfo {author} {\bibfnamefont {A.~V.}\ \bibnamefont
  {Shevelkov}},\ }\href
  {http://link.springer.com/article/10.1023/A%3A1015010907973} {\bibfield
  {journal} {\bibinfo  {journal} {Russ. Chem. Bull. Int. Ed.}\ }\textbf
  {\bibinfo {volume} {50}},\ \bibinfo {pages} {2304} (\bibinfo {year}
  {2001})}\BibitemShut {NoStop}%
\bibitem [{\citenamefont {Filatova}\ \emph {et~al.}(2007)\citenamefont
  {Filatova}, \citenamefont {Gurin}, \citenamefont {Kloo}, \citenamefont
  {Kulbachinskii}, \citenamefont {Kuznetsov}, \citenamefont {Kytin},
  \citenamefont {Lindsjo},\ and\ \citenamefont {Popovkin}}]{Filatova2007}%
  \BibitemOpen
  \bibfield  {author} {\bibinfo {author} {\bibfnamefont {T.~G.}\ \bibnamefont
  {Filatova}}, \bibinfo {author} {\bibfnamefont {P.~V.}\ \bibnamefont {Gurin}},
  \bibinfo {author} {\bibfnamefont {L.}~\bibnamefont {Kloo}}, \bibinfo {author}
  {\bibfnamefont {V.~A.}\ \bibnamefont {Kulbachinskii}}, \bibinfo {author}
  {\bibfnamefont {A.~N.}\ \bibnamefont {Kuznetsov}}, \bibinfo {author}
  {\bibfnamefont {V.~G.}\ \bibnamefont {Kytin}}, \bibinfo {author}
  {\bibfnamefont {M.}~\bibnamefont {Lindsjo}}, \ and\ \bibinfo {author}
  {\bibfnamefont {B.~A.}\ \bibnamefont {Popovkin}},\ }\href {\doibase
  10.1016/j.jssc.2007.01.010} {\bibfield  {journal} {\bibinfo  {journal} {J.
  Solid State Chem.}\ }\textbf {\bibinfo {volume} {180}},\ \bibinfo {pages}
  {1103} (\bibinfo {year} {2007})}\BibitemShut {NoStop}%
\bibitem [{\citenamefont {Mak}\ \emph {et~al.}(2010)\citenamefont {Mak},
  \citenamefont {Lee}, \citenamefont {Hone}, \citenamefont {Shan},\ and\
  \citenamefont {Heinz}}]{Mak2010}%
  \BibitemOpen
  \bibfield  {author} {\bibinfo {author} {\bibfnamefont {K.~F.}\ \bibnamefont
  {Mak}}, \bibinfo {author} {\bibfnamefont {C.}~\bibnamefont {Lee}}, \bibinfo
  {author} {\bibfnamefont {J.}~\bibnamefont {Hone}}, \bibinfo {author}
  {\bibfnamefont {J.}~\bibnamefont {Shan}}, \ and\ \bibinfo {author}
  {\bibfnamefont {T.~F.}\ \bibnamefont {Heinz}},\ }\href {\doibase
  10.1103/PhysRevLett.105.136805} {\bibfield  {journal} {\bibinfo  {journal}
  {Phys. Rev. Lett.}\ }\textbf {\bibinfo {volume} {105}},\ \bibinfo {pages}
  {136805} (\bibinfo {year} {2010})}\BibitemShut {NoStop}%
\bibitem [{\citenamefont {Heyd}\ \emph {et~al.}(2005)\citenamefont {Heyd},
  \citenamefont {Peralta}, \citenamefont {Scuseria},\ and\ \citenamefont
  {Martin}}]{Heyd2005}%
  \BibitemOpen
  \bibfield  {author} {\bibinfo {author} {\bibfnamefont {J.}~\bibnamefont
  {Heyd}}, \bibinfo {author} {\bibfnamefont {J.~E.}\ \bibnamefont {Peralta}},
  \bibinfo {author} {\bibfnamefont {G.~E.}\ \bibnamefont {Scuseria}}, \ and\
  \bibinfo {author} {\bibfnamefont {R.~L.}\ \bibnamefont {Martin}},\ }\href
  {\doibase http://dx.doi.org/10.1063/1.2085170} {\bibfield  {journal}
  {\bibinfo  {journal} {The Journal of Chemical Physics}\ }\textbf {\bibinfo
  {volume} {123}},\ \bibinfo {eid} {174101} (\bibinfo {year}
  {2005})}\BibitemShut {NoStop}%
\bibitem [{\citenamefont {Liu}\ \emph {et~al.}(2014)\citenamefont {Liu},
  \citenamefont {Neal}, \citenamefont {Zhu}, \citenamefont {Luo}, \citenamefont
  {Xu}, \citenamefont {Tománek},\ and\ \citenamefont {Ye}}]{Liu2014}%
  \BibitemOpen
  \bibfield  {author} {\bibinfo {author} {\bibfnamefont {H.}~\bibnamefont
  {Liu}}, \bibinfo {author} {\bibfnamefont {A.~T.}\ \bibnamefont {Neal}},
  \bibinfo {author} {\bibfnamefont {Z.}~\bibnamefont {Zhu}}, \bibinfo {author}
  {\bibfnamefont {Z.}~\bibnamefont {Luo}}, \bibinfo {author} {\bibfnamefont
  {X.}~\bibnamefont {Xu}}, \bibinfo {author} {\bibfnamefont {D.}~\bibnamefont
  {Tománek}}, \ and\ \bibinfo {author} {\bibfnamefont {P.~D.}\ \bibnamefont
  {Ye}},\ }\href {\doibase 10.1021/nn501226z} {\bibfield  {journal} {\bibinfo
  {journal} {ACS Nano}\ }\textbf {\bibinfo {volume} {8}},\ \bibinfo {pages}
  {4033} (\bibinfo {year} {2014})}\BibitemShut {NoStop}%
\bibitem [{\citenamefont {Bernevig}, \citenamefont {Hughes},\ and\
  \citenamefont {Zhang}(2006)}]{Bernevig2006}%
  \BibitemOpen
  \bibfield  {author} {\bibinfo {author} {\bibfnamefont {B.~A.}\ \bibnamefont
  {Bernevig}}, \bibinfo {author} {\bibfnamefont {T.~L.}\ \bibnamefont
  {Hughes}}, \ and\ \bibinfo {author} {\bibfnamefont {S.-C.}\ \bibnamefont
  {Zhang}},\ }\href {\doibase 10.1126/science.1133734} {\bibfield  {journal}
  {\bibinfo  {journal} {Science}\ }\textbf {\bibinfo {volume} {314}},\ \bibinfo
  {pages} {1757} (\bibinfo {year} {2006})}\BibitemShut {NoStop}%
\bibitem [{\citenamefont {Zhang}\ \emph {et~al.}(2009)\citenamefont {Zhang},
  \citenamefont {Liu}, \citenamefont {Qi}, \citenamefont {Dai}, \citenamefont
  {Fang},\ and\ \citenamefont {Zhang}}]{Zhang2009}%
  \BibitemOpen
  \bibfield  {author} {\bibinfo {author} {\bibfnamefont {H.}~\bibnamefont
  {Zhang}}, \bibinfo {author} {\bibfnamefont {C.-X.}\ \bibnamefont {Liu}},
  \bibinfo {author} {\bibfnamefont {X.-L.}\ \bibnamefont {Qi}}, \bibinfo
  {author} {\bibfnamefont {X.}~\bibnamefont {Dai}}, \bibinfo {author}
  {\bibfnamefont {Z.}~\bibnamefont {Fang}}, \ and\ \bibinfo {author}
  {\bibfnamefont {S.-C.}\ \bibnamefont {Zhang}},\ }\href {\doibase
  10.1038/nphys1270} {\bibfield  {journal} {\bibinfo  {journal} {Nat Phys}\
  }\textbf {\bibinfo {volume} {5}},\ \bibinfo {pages} {438} (\bibinfo {year}
  {2009})}\BibitemShut {NoStop}%
\bibitem [{\citenamefont {Sabater}\ \emph {et~al.}(2013)\citenamefont
  {Sabater}, \citenamefont {Gos{\'a}lbez-Mart{\'\i}nez}, \citenamefont
  {Fern{\'a}ndez-Rossier}, \citenamefont {Rodrigo}, \citenamefont {Untiedt},\
  and\ \citenamefont {Palacios}}]{Sabater2013}%
  \BibitemOpen
  \bibfield  {author} {\bibinfo {author} {\bibfnamefont {C.}~\bibnamefont
  {Sabater}}, \bibinfo {author} {\bibfnamefont {D.}~\bibnamefont
  {Gos{\'a}lbez-Mart{\'\i}nez}}, \bibinfo {author} {\bibfnamefont
  {J.}~\bibnamefont {Fern{\'a}ndez-Rossier}}, \bibinfo {author} {\bibfnamefont
  {J.~G.}\ \bibnamefont {Rodrigo}}, \bibinfo {author} {\bibfnamefont
  {C.}~\bibnamefont {Untiedt}}, \ and\ \bibinfo {author} {\bibfnamefont
  {J.~J.}\ \bibnamefont {Palacios}},\ }\href {\doibase
  10.1103/PhysRevLett.110.176802} {\bibfield  {journal} {\bibinfo  {journal}
  {Phys. Rev. Lett.}\ }\textbf {\bibinfo {volume} {110}},\ \bibinfo {pages}
  {176802} (\bibinfo {year} {2013})}\BibitemShut {NoStop}%
\end{thebibliography}%

\end{document}